\documentclass[aps,twocolumn,reprint]{revtex4-2}
\usepackage{graphicx}
\usepackage{amsmath}

\begin{document}

\title{Propagation of Dirac waves through various temporal interfaces, slabs, and crystals}

\author{Seulong Kim}
\affiliation{Research Institute of Basic Sciences, Ajou University,
Suwon 16499, Korea}
\author{Kihong Kim}
\email{khkim@ajou.ac.kr}
\affiliation{Department of Physics, Ajou University, Suwon 16499, Korea}
\affiliation{School of Physics, Korea Institute for Advanced Study, Seoul 02455, Korea}
\begin{abstract}
We investigate the influence of the temporal variations of various medium parameters on the propagation of Dirac-type waves in
materials where the quasiparticles are described by a generalized version of the pseudospin-1/2 Dirac equation.
Our considerations also include the propagation of electromagnetic waves in metamaterials with the Dirac-type dispersion. We focus on the variations of the scalar and vector potentials, mass,
Fermi velocity, and tilt velocity describing the Dirac cone tilt.
We derive the scattering coefficients associated with the temporal interfaces and slabs analytically and
find that the temporal scattering is caused by the changes of the mass, Fermi velocity, and vector potential, but does not arise from the changes of the scalar potential and tilt velocity.
We also explore the conditions under which the temporal Brewster effect and total interband transition occur and calculate the change in total wave energy.
We examine bilayer Dirac temporal crystals where parameters switch between two different sets of values periodically and prove that these systems do not have momentum gaps.
Finally, we assess the potential for observing these temporal scattering effects in experiments.
\end{abstract}

\maketitle

\section{Introduction}

The propagation of waves in time-varying media where the medium parameters vary as a function of time
is an old topic that has been studied extensively in many branches of physics and engineering \cite{Mor1958,Felsen1970,Fante1971,Kalluri2010}.
Since quantum particles can be described by wave equations such as the Schr\"odinger,
Dirac, and Klein-Gordon equations in many physical situations, this topic is also relevant for the study of quantum materials
as well as of the propagation of classical waves.
Recently, there has been a strong renewed interest in the propagation of electromagnetic waves in
time-varying dielectric media, where the temporal variation of the dielectric permittivity or other parameters
causes various temporal scattering effects such as the temporal reflection and refraction \cite{Mendonca2002,Agrawal2014,Hay2016,Caloz2020a,Caloz2020b,Ramaccia2020,Boyd2020,Galiffi2022}.
The temporal reflection refers to the appearance of backward propagating waves due to a temporal variation of the medium parameters.
Many other interesting phenomena including temporal circular birefringence, temporal aiming, and temporal Brewster angle have been proposed to arise
in general anisotropic and bianisotropic media \cite{Zhang2015,Wang2020,Li2022,Engheta2020,Engheta2021}. In the presence of periodic temporal variations,
it has been long known that there appear momentum gaps (or $k$ gaps), which are analogous to the frequency gaps appearing in
spatially periodic photonic crystals \cite{Bian2007,Sanchez2009,Kout2018,Segev2018,Galiffi2020}. Wave propagation in such photonic temporal crystals has been studied in the framework of
the Floquet analysis \cite{Oka2019}.

In contrast to the extensive research on the photonics of time-varying media,
there have been far fewer studies on similar problems for quantum wave equations.
In the case of the Schr\"odinger equation, the temporal scattering does not occur because it is a first-order differential equation
in time. However, the temporal scattering does occur in the systems governed by relativistic wave equations such as the Dirac and Klein-Gordon equations.
For the Dirac equation in two dimensions, the temporal scattering effect due to the variation of the mass term
has been studied recently in the context of quantum time mirrors \cite{Reck2017,Reck2018,Junk2020}.
In this paper, we study the propagation of waves governed by a generalized form of the pseudospin-1/2
Dirac equation in the presence of various kinds of time-varying perturbations.
More specifically, we consider the temporal variations of the scalar and vector potentials, Fermi velocity, tilt velocity describing
the magnitude and the direction of the Dirac cone tilt, and mass describing the band gap between the upper and lower Dirac cones.
For the simplest configurations such as temporal interfaces and slabs,
we derive the analytical expressions of the
temporal scattering coefficients and prove that the variations of the vector potential, Fermi velocity,
and mass cause the temporal scattering effects, whereas those of the scalar potential and tilt velocity do not.
We also derive analytically the explicit conditions for the temporal equivalents of the total transmission
and the total interband transition. The temporal total transmission may also be called temporal Brewster effect \cite{Engheta2021}. In addition, we derive the expressions for the change of the total wave energy
for temporal interfaces and slabs. In the case of bilayer temporal crystals
where the medium parameters alternate between two different values periodically in time, we prove that the momentum gaps never appear in the systems satisfying
the pseudospin-1/2 Dirac equation. This is in a sharp contrast to the case of electromagnetic waves and other classical
waves.

There exist many two- and three-dimensional materials where the quasiparticles satisfy the generalized pseudospin-1/2
Dirac equation \cite{Castro2009,Weh2014,Zhang2009,Feng2017}. In such materials, temporal variations of the vector potential, Fermi velocity, and mass
can be readily realized experimentally and the consequences of the temporal scattering will be manifested in various transport properties.
On the other hand, it is possible to fabricate electromagnetic and elastic metamaterials mimicking the Dirac materials where
the dispersion relations are of the Dirac type \cite{Zhang2012,Garreau2017,Mili2019,Li2021,Zhang2022}. The phenomena investigated in the present study can also be tested in experiments
involving such metamaterials.

The rest of this paper is organized as follows.
In Sec.~\ref{sec2}, we introduce a generalized version of the pseudospin-1/2 Dirac equation in two dimesnions.
In Sec.~\ref{sec3}, we derive the temporal scattering coefficients for temporal interfaces and the conditions
for the temporal Brewster effect and total interband transition analytically. We also derive the
expressions for the change of the total wave energy.
Similar calculations are performed for temporal slabs in Sec.~\ref{sec4}.
In Sec.~\ref{sec5}, we derive the dispersion relations for general bilayer Dirac temporal crystals and prove that
there never appears a momentum gap in such systems.
Finally, we give a brief discussion of the experimental feasibility of the effects considered in this work
and conclude the paper in Sec.~\ref{sec6}.

\begin{figure}
\includegraphics[width=7.5cm]{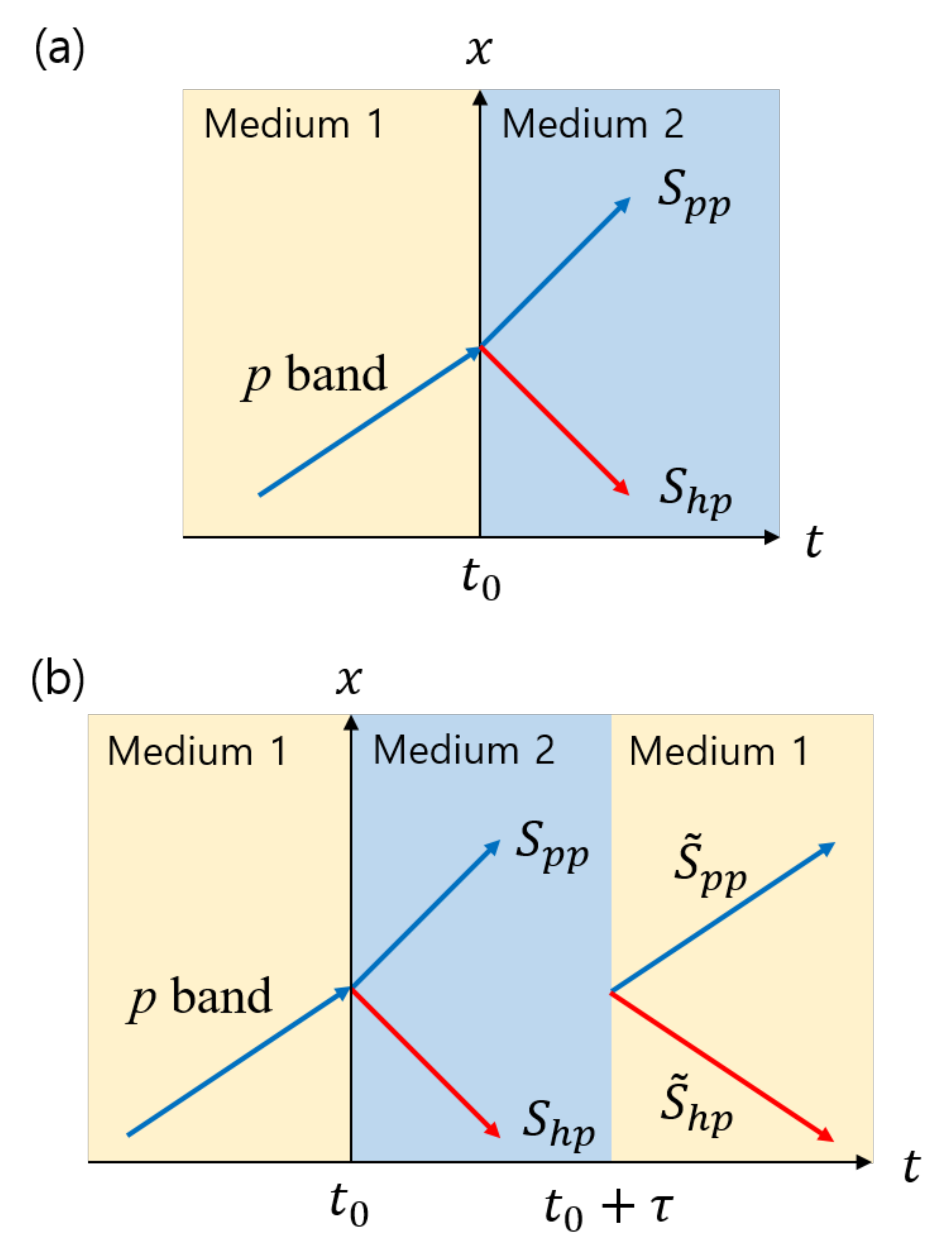}
\caption{Schematics of the temporal scattering processes of waves satisfying the generalized Dirac equation
through (a) a temporal interface at $t=t_0$ and (b) a temporal slab of interval $\tau$,
when a $p$-band wave propagates initially in a uniform medium.
The blue and red arrows denote the group velocities of $p$-
and $h$-band waves, respectively. In the absence of Dirac cone tilt, the two group velocities
are directed in precisely opposite directions. $S_{pp}$ and $S_{hp}$ denote the intraband and interband scattering coefficients
representing $p\rightarrow p$ and  $p\rightarrow h$ transitions at the interface at $t=t_0$, respectively.
In (b), $\tilde S_{pp}$ and $\tilde S_{hp}$ denote the scattering coefficients for
the temporal slab of interval $\tau$.}
\label{fig1}
\end{figure}

\section{Wave equation}
\label{sec2}

We consider a generalized form of the effective Hamiltonian for massive pseudospin-1/2 Dirac particles moving in the two-dimensional (2D) $xy$ plane given by
\begin{widetext}
\begin{eqnarray}
{\mathcal H}&=&v_{x}\sigma_x\pi_x+v_{y}\sigma_y\pi_y+\left(v_{tx}\pi_x +v_{ty}\pi_y\right) I
+UI+M\sigma_z\nonumber\\
&=&\begin{pmatrix} v_{tx}\pi_x+v_{ty}\pi_y+U+M & v_x\pi_x-iv_y\pi_y \\ v_x\pi_x+iv_y\pi_y & v_{tx}\pi_x+v_{ty}\pi_y+U-M \end{pmatrix},
\label{eq:ham0}
\end{eqnarray}
\end{widetext}
where
\begin{eqnarray}
\pi_x=\hbar k_x+eA_x,~~\pi_y=\hbar k_y+eA_y.
\label{eq:ham1}
\end{eqnarray}
The parameters $v_x$ and $v_y$ are the anisotropic Fermi velocity components and $v_{tx}$ and $v_{ty}$ are the $x$ and $y$ components of the
tilt velocity describing the direction and the magnitude of the Dirac cone tilt in the momentum space \cite{Been2016,Nguyen2018}. $U$ is the scalar
potential and $A_x$ and $A_y$ are the $x$ and $y$ components of the vector potential. $M$ is the mass energy describing
the energy gap between the upper and lower Dirac cones and $e$ is the electron charge. $k_x$ and
$k_y$ are the components of the wave vector and $\sigma_x$, $\sigma_y$, and $\sigma_z$ are the Pauli matrices. $I$ is the $2\times 2$ unity matrix.

In this paper, we consider the situation where one or several of
the parameters $U$, $A_x$, $A_y$, $v_x$, $v_y$, $v_{tx}$, $v_{ty}$, and $M$ are functions of time,
while being uniform in the entire space. Then the wave-vector components $k_x$ and $k_y$ are constants of the motion.
The time-dependent Dirac equation in two dimensions for the two-component vector wave function $\Psi$ [$=\left( \psi_1, \psi_2\right)^{\rm T}$] is
\begin{eqnarray}
i\hbar \frac{d\Psi}{dt}={\mathcal H}\Psi.
\end{eqnarray}
If all the parameters are constants independent of time, then we can easily solve this equation
by assuming that the wave function depends on time
as $e^{-i\omega t}$. The dispersion relation that follows from this takes the form
\begin{eqnarray}
\omega={\bf v}_t\cdot{\bf q}+\frac{U}{\hbar}\pm\Omega,
\end{eqnarray}
where
\begin{eqnarray}
&&\Omega=\sqrt{\mu^2+\vert\nu\vert^2},~\mu=\frac{M}{\hbar},~\nu=v_x q_x-iv_y q_y,\nonumber\\
&&{\bf q}={\bf k}+\frac{e{\bf A}}{\hbar}.
\end{eqnarray}
${\bf v}_t$, $\bf q$, $\bf k$, and $\bf A$ are 2D vectors with the $x$ and $y$ components.
The two solutions with the plus and minus signs represent respectively the particle-like and hole-like bands corresponding to the upper and lower Dirac cones.
The group velocities for the two bands, which we call $p$ and $h$ bands respectively, are given by
\begin{eqnarray}
&&{\bf v}_g=(v_{gx},v_{gy})^{\rm T}=\left(\frac{\partial \omega}{\partial k_x},\frac{\partial \omega}{\partial k_y}\right)^{\rm T} \nonumber\\
&&~~~=\left\{\begin{array}{l l}
{\bf v}_t+\frac{1}{\Omega}\left({v_x}^2q_x,{v_y}^2q_y\right)^{\rm T}, & \quad \mbox{{\it p} band}\\
{\bf v}_t-\frac{1}{\Omega}\left({v_x}^2q_x,{v_y}^2q_y\right)^{\rm T}, & \quad \mbox{{\it h} band}
\end{array}\right..
\end{eqnarray}
When the tilt velocity ${\bf v}_t$ or the vector potential $\bf A$ is nonzero, or when the Fermi velocity is anisotropic
such that $v_x\ne v_y$,
the group velocity becomes anisotropic. In the absence of the tilt, the group velocities for the $p$ and $h$ bands are directed precisely opposite to each other. We notice that ${\bf v}_g$ depends on ${\bf v}_t$, $\bf v$ [$=(v_x,v_y)^{\rm T}$], $\bf A$, and $M$, but is
independent of $U$.

In the stationary regime, the two components of the wave function, $\psi_1$ and $\psi_2$, are proportional to each other so that
\begin{eqnarray}
\psi_2=\left\{\begin{array}{l l} \chi_p\psi_1, & \quad \mbox{{\it p} band}\\
\chi_h\psi_1, & \quad \mbox{{\it h} band}\end{array}\right.,
\end{eqnarray}
where
\begin{eqnarray}
&&\chi_p=\frac{\Omega-\mu}{\nu},~
\chi_h=-\frac{\Omega+\mu}{\nu}.
\end{eqnarray}
The quantities $\chi_p$ and $\chi_h$ can be considered as the effective wave impedances for $p$- and $h$-band waves respectively.

\begin{figure}
\includegraphics[width=8cm]{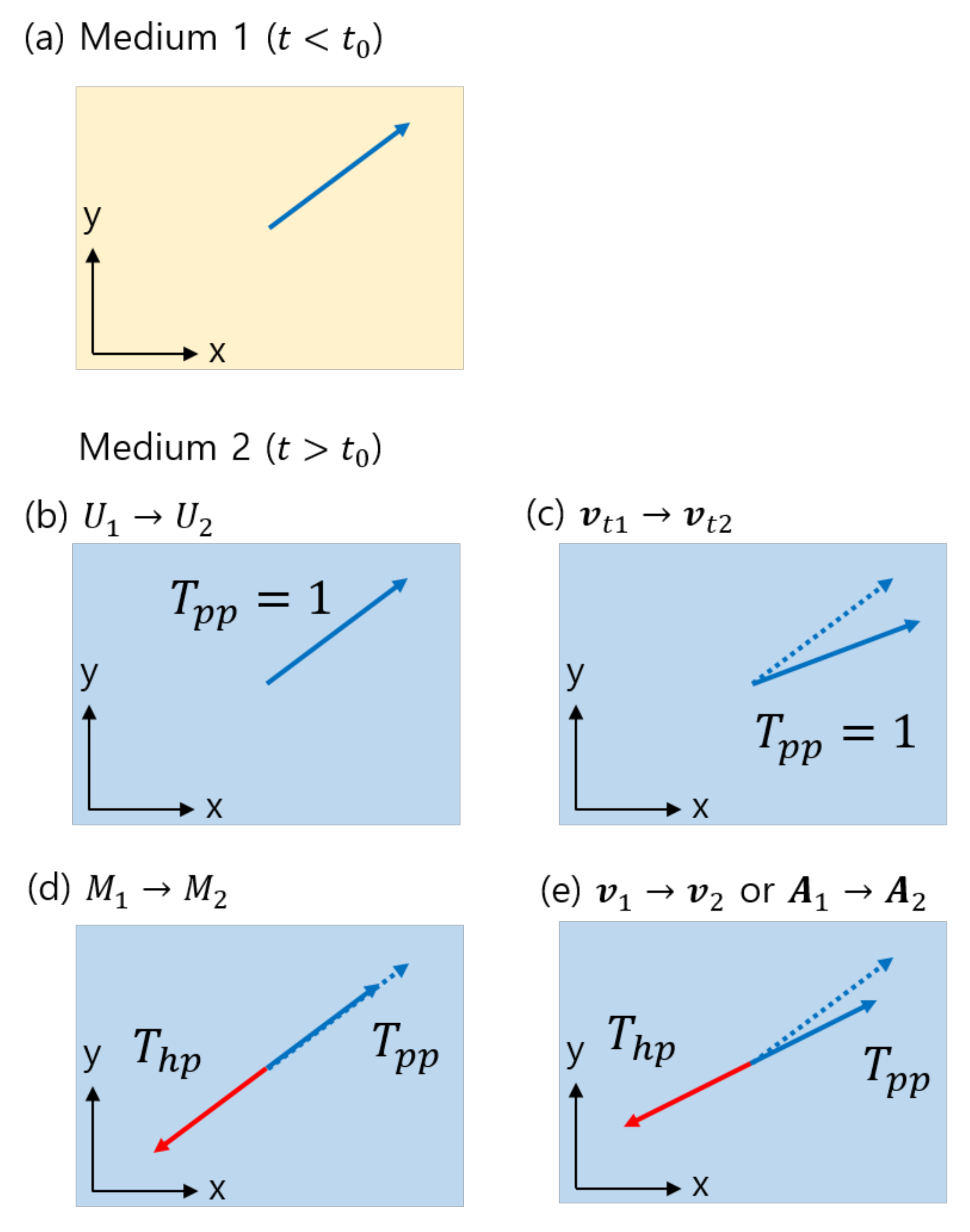}
\caption{Schematics of the temporal scattering of (a) a $p$-band wave propagating in a uniform medium due to
sudden changes of (b) the scalar potential, (c) tilt velocity, (d) mass, (e) Fermi velocity  or
vector potential. The blue and red arrows denote the group velocities of $p$-
and $h$-band waves, respectively. The dotted and straight arrows in (b-e) represent the initial and scattered waves, respectively.
In (b), there is neither the scattered $h$-band wave nor the change of the group velocity.
In (c), there is no scattered $h$-band wave, but the group velocity is changed.
In (d), in the absence of Dirac cone tilt, the group velocities of the scattered waves are the same as or opposite to that of the initial wave. In (e), the group velocities of the scattered waves are generally not parallel
or antiparallel to that of the initial wave.}
\label{fig2}
\end{figure}

\begin{figure}
\includegraphics[width=8.6cm]{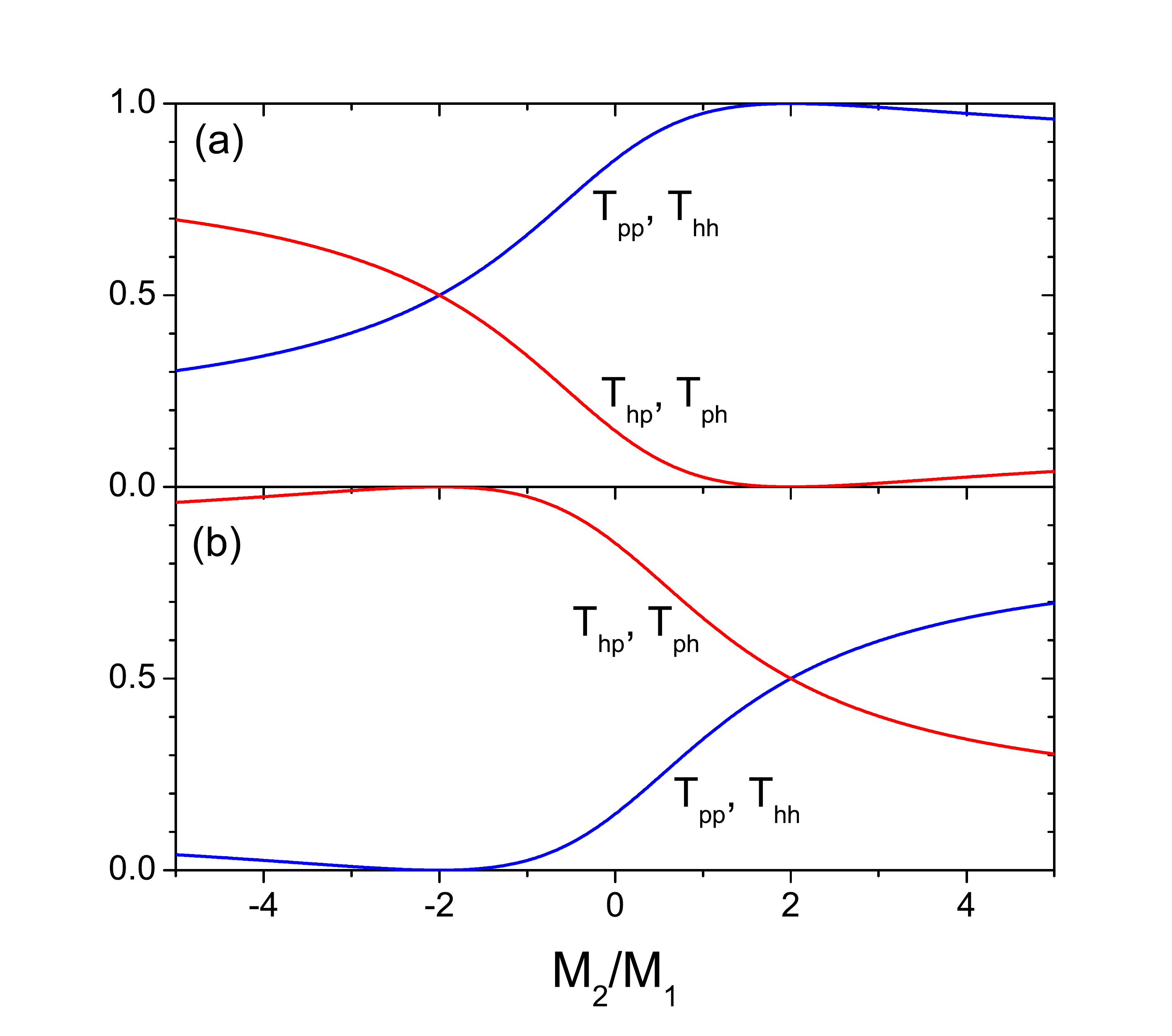}
\caption{Transmittances ($T_{pp}=T_{hh}$) and interband transition rates ($T_{hp}=T_{ph}$)
for a temporal interface plotted versus the mass ratio $M_2/M_1$. The sums $T_{pp}+T_{hp}$
and $T_{hh}+T_{ph}$ are always equal to 1. The common values of the parameters are $k_y=A_{y1}=A_{y2}=0$ and
$M_1/(\hbar k_x v_{x1})=1$. In (a), we take $v_{x2}q_{x2}=2v_{x1}q_{x1}$. When $M_2/M_1=2$, the temporal Brewster effect for which
$T_{pp}=T_{hh}=1$ arises. In (b), we take $v_{x2}q_{x2}=-2v_{x1}q_{x1}$. When $M_2/M_1=-2$, the temporal total interband transition for which
$T_{pp}=T_{hh}=0$ arises.}
\label{fig3}
\end{figure}

\section{Temporal interface}
\label{sec3}

\subsection{Temporal scattering coefficients}

We assume that, at $t=t_0$, the parameters $U$, $A_x$, $A_y$, $v_x$, $v_y$, $v_{tx}$, $v_{ty}$, and $M$ change from
$U_1$, $A_{x1}$, $A_{y1}$, $v_{x1}$, $v_{y1}$, $v_{tx1}$, $v_{ty1}$, and $M_1$ to
$U_2$, $A_{x2}$, $A_{y2}$, $v_{x2}$, $v_{y2}$, $v_{tx2}$, $v_{ty2}$, and $M_2$ abruptly.
Let us first suppose that the state before the change is a $p$-band state. Then the wave functions before and after the temporal change can be written as
\begin{widetext}
\begin{eqnarray}
\Psi=\left\{\begin{array}{l l}
\begin{pmatrix} 1\\ \chi_{p1}\end{pmatrix} e^{-i\omega_{p1} (t-t_0)}, & \quad \mbox{if $t<t_0$ }\\
S_{pp}\begin{pmatrix} 1\\ \chi_{p2}\end{pmatrix} e^{-i\omega_{p2} (t-t_0)}+S_{hp}\begin{pmatrix} 1\\ \chi_{h2}\end{pmatrix} e^{-i\omega_{h2} (t-t_0)}, & \quad \mbox{if $t\ge t_0$ }
\end{array}\right.,
\end{eqnarray}
\end{widetext}
where
$\omega_{pj}$, $\omega_{hj}$, $\chi_{pj}$, and $\chi_{hj}$ ($j=1,2$) are defined by
\begin{eqnarray}
&&\omega_{pj}=v_{txj}q_{xj}+v_{tyj}q_{yj}+\frac{U_j}{\hbar}+\Omega_j,\nonumber\\
&&\omega_{hj}=v_{txj}q_{xj}+v_{tyj}q_{yj}+\frac{U_j}{\hbar}-\Omega_j,\nonumber\\
&&\chi_{pj}=\frac{\Omega_j-\mu_j}{\nu_j},~
\chi_{hj}=-\frac{\Omega_j+\mu_j}{\nu_j},\nonumber\\
&&\Omega_j=\sqrt{{\mu_j}^2+{\vert\nu_j\vert}^2},~\mu_j=\frac{M_j}{\hbar},~\nu_j=v_{xj}q_{xj}-iv_{yj}q_{yj},\nonumber\\
&&q_{xj}=k_x+\frac{eA_{xj}}{\hbar},~q_{yj}=k_y+\frac{eA_{yj}}{\hbar},
\end{eqnarray}
and $S_{pp}$ and $S_{hp}$ are the intraband and interband scattering coefficients representing $p\rightarrow p$ and
$p\rightarrow h$ transitions, respectively. From the continuity of $\Psi$ at the temporal interface at $t=t_0$,
we obtain
\begin{eqnarray}
&&S_{pp}=\frac{\chi_{p1}-\chi_{h2}}{\chi_{p2}-\chi_{h2}}=\frac{\nu_2}{2\Omega_2}\left(\frac{\Omega_1-\mu_1}{\nu_1}+\frac{\Omega_2+\mu_2}{\nu_2}\right),\nonumber\\
&&S_{hp}=\frac{\chi_{p2}-\chi_{p1}}{\chi_{p2}-\chi_{h2}}=\frac{\nu_2}{2\Omega_2}\left(-\frac{\Omega_1-\mu_1}{\nu_1}+\frac{\Omega_2-\mu_2}{\nu_2}\right).\nonumber\\
\label{eq:sc1}
\end{eqnarray}
Similarly, when the state before the temporal change is a $h$-band state, the wave functions can be written as
\begin{widetext}
\begin{eqnarray}
\Psi=\left\{\begin{array}{l l}
\begin{pmatrix} 1\\ \chi_{h1}\end{pmatrix} e^{-i\omega_{h1} (t-t_0)}, & \quad \mbox{if $t<t_0$ }\\
S_{hh}\begin{pmatrix} 1\\ \chi_{h2}\end{pmatrix} e^{-i\omega_{h2} (t-t_0)}+S_{ph}\begin{pmatrix} 1\\ \chi_{p2}\end{pmatrix} e^{-i\omega_{p2} (t-t_0)}, & \quad \mbox{if $t\ge t_0$ }
\end{array}\right.,
\end{eqnarray}
\end{widetext}
where the scattering coefficients $S_{hh}$ and $S_{ph}$ representing $h\rightarrow h$ and
$h\rightarrow p$ transitions respectively are given by
\begin{eqnarray}
&&S_{hh}=\frac{\chi_{p2}-\chi_{h1}}{\chi_{p2}-\chi_{h2}}=\frac{\nu_2}{2\Omega_2}\left(\frac{\Omega_1+\mu_1}{\nu_1}+\frac{\Omega_2-\mu_2}{\nu_2}\right),\nonumber\\
&&S_{ph}=\frac{\chi_{h1}-\chi_{h2}}{\chi_{p2}-\chi_{h2}}=\frac{\nu_2}{2\Omega_2}\left(-\frac{\Omega_1+\mu_1}{\nu_1}+\frac{\Omega_2+\mu_2}{\nu_2}\right).\nonumber\\
\label{eq:sc2}
\end{eqnarray}
The scattering coefficients $S_{pp}$, $S_{hp}$, $S_{hh}$, and $S_{ph}$ depend on $\Omega_j$, $\mu_j$, and $\nu_j$
($j=1,2$), which in turn depend on $M_j$, ${\bf v}_{j}$, and ${\bf A}_{j}$, but not on $U_j$ and ${\bf v}_{tj}$.
In other words, temporal variations of the mass, Fermi velocity, and vector potential cause
temporal scattering, but those of the scalar potential and tilt velocity do not.

In Fig.~\ref{fig1}(a), we show a schematic of the temporal scattering process through a temporal interface at $t=t_0$,
where the blue and red arrows denote the group velocities of $p$-
and $h$-band waves, respectively. In the absence of Dirac cone tilt, the two group velocities
are directed in precisely opposite directions.

In Fig.~\ref{fig2}, we show schematics of the temporal scattering due to sudden changes of $U$, ${\bf v}_t$, $M$, $\bf v$, and $\bf A$.
When the scalar potential is varied, there appears neither the scattered wave nor the change of the group velocity.
When the tilt velocity is varied, there appears no scattered wave, but the group velocity is changed. When the mass is varied
in the absence of Dirac cone tilt, the group velocities of the scattered waves are the same as or opposite to that of the initial wave.
When the Fermi velocity or the vector potential is varied, the group velocities of the scattered waves are generally not parallel or antiparallel to that of the initial wave. If the mass is zero all the time in this case, some special situations can occur as is explained in the next subsection.

\begin{figure}
\includegraphics[width=8.6cm]{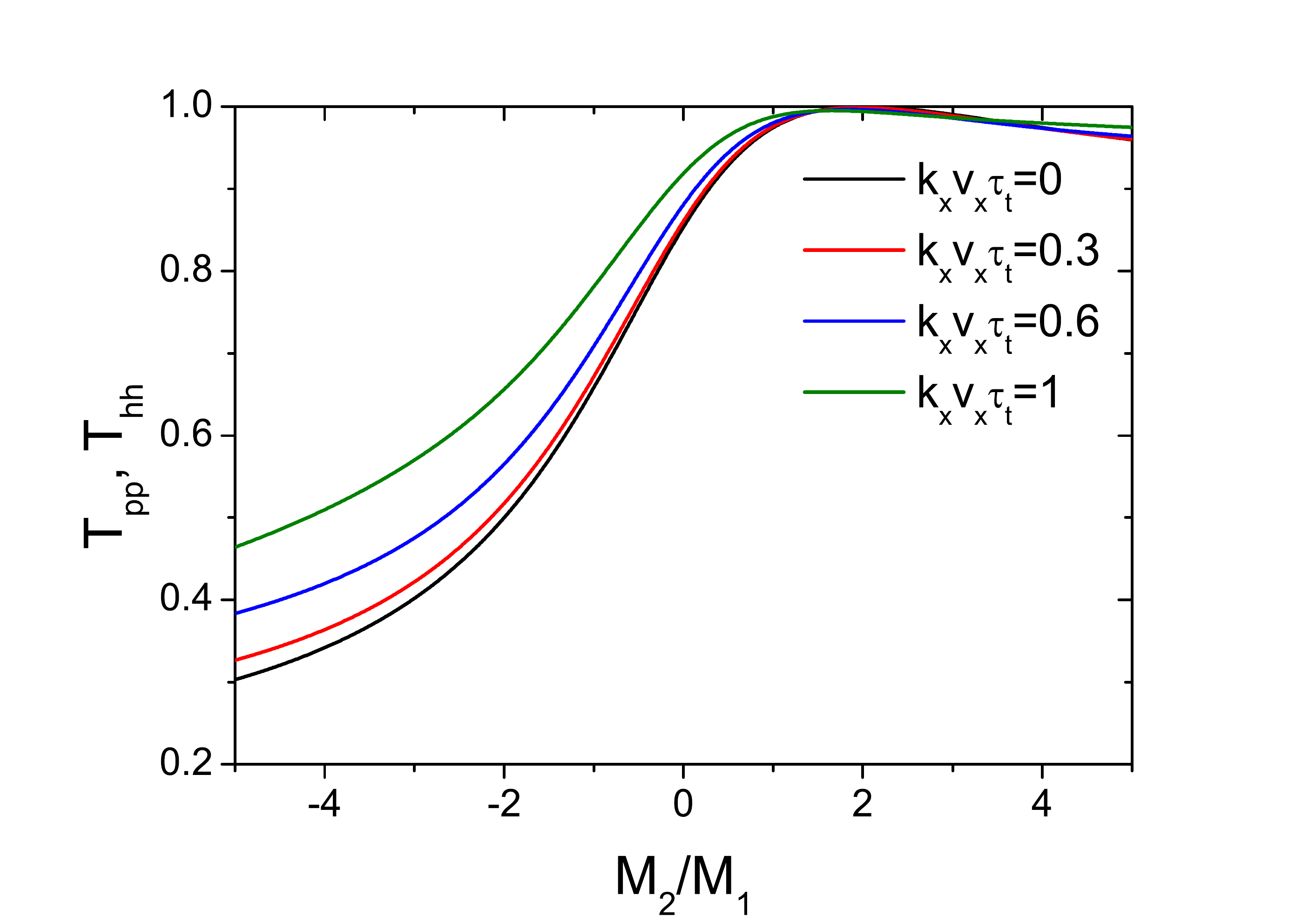}
\caption{Transmittances ($T_{pp}=T_{hh}$) at the temporal interface corresponding to
Fig.~\ref{fig3}(a) compared to those
obtained for continuous interfaces with various values of the interval $\tau_t$.
It is assumed that $k_y=A_{y}=0$ and $v_x$ is constant. Before $t=0$, $M=M_1=\hbar k_x v_{x}$
and $A_x=0$. During the interval $0<t<\tau_t$, $M$
and the normalized vector potential $a_{x}$ $[= eA_{x}/(\hbar k_x)]$ change according to
$M=M_1+(M_2-M_1)t/\tau_t$ and $a_x=(t/\tau_t)^2$, respectively. After $t=\tau_t$, they remain to be constants $M=M_2$ and $a_x=1$. With increasing $\tau_t$, the maximum transmittance shifts to smaller $M_2/M_1$ ratios, while its value remains nearly 1.
}
\label{ff4}
\end{figure}

\begin{figure}
\includegraphics[width=8.6cm]{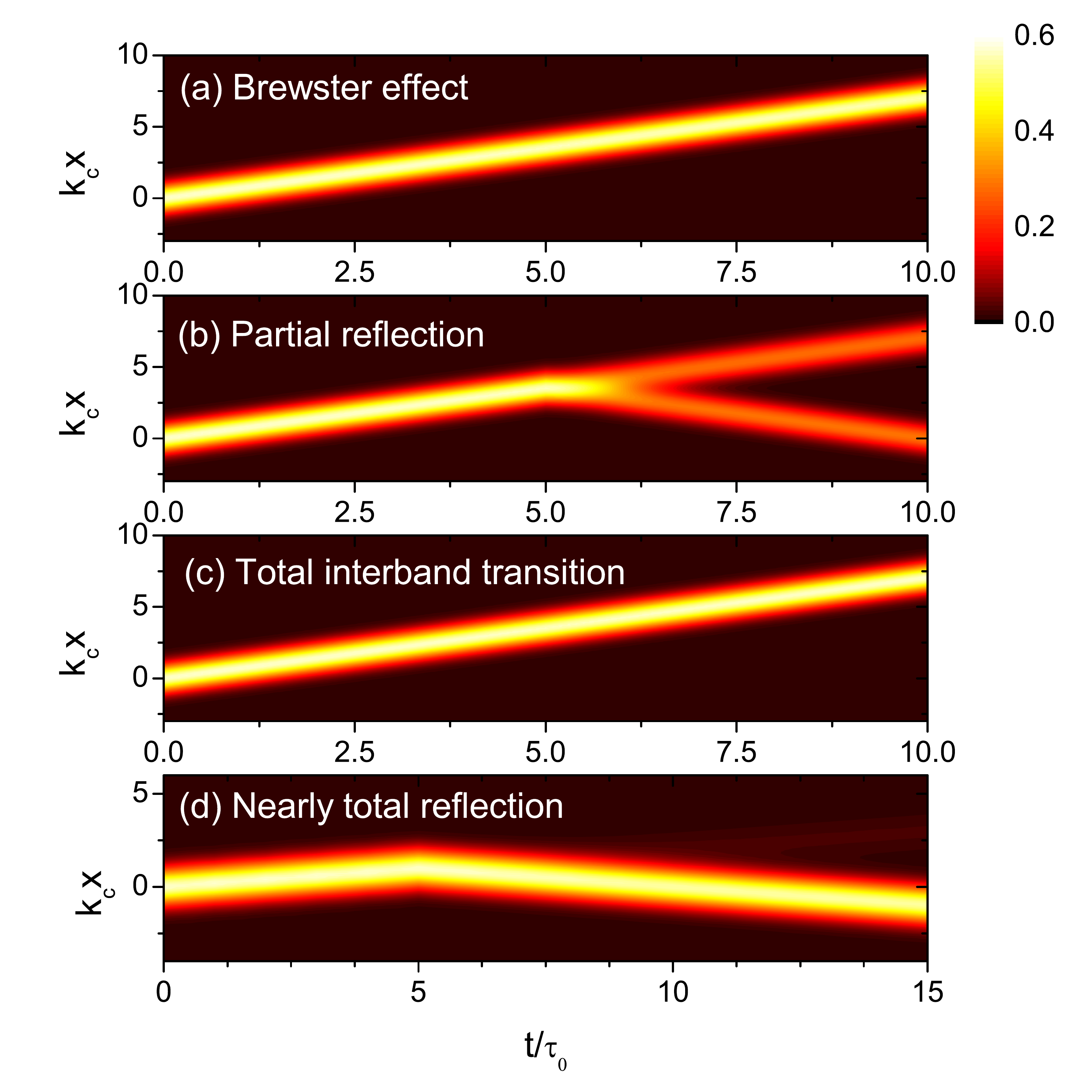}
\caption{Propagation of a Gaussian pulse
through a sudden temporal interface occurring at $t=5\tau_0$.
It is assumed that $k_y=A_{y}=0$ and $v_x$ is constant.
At $t=0$, the pulse whose central wave number is $k_c$ is located at $x=0$ and propagates towards the $+x$ direction. At $t=5\tau_0$, the mass energy $M$ and the normalized vector potential $a_{x}$ $[= eA_{x}/(\hbar k_c)]$ change from $M_1=\hbar k_{c} v_{x}$
and $a_{x1}=0$ to (a) $M_2=2M_1$ and $a_{x2}=1$, (b) $M_2=-2M_1$ and $a_{x2}=1$, (c) $M_2=-2M_1$ and $a_{x2}=-3$.
In (d), the mass changes from $5\hbar k_{c} v_{x}$ to $-5\hbar k_{c} v_{x}$, while $A_x$ remains to be zero.
The pulse width is characterized by the parameter $\sigma_k$ ($=0.1k_c$) and the time unit is defined by
$\tau_0=1/(\sigma_k v_x)$.
In (b) and (d), $50 \%$ and about $96.1 \%$ of the pulse are reflected in the opposite direction, respectively. }
\label{ff5}
\end{figure}

\subsection{Temporal Brewster effect and temporal total interband transition}

Let us suppose that the relationships
\begin{eqnarray}
&&v_{x1}q_{x1}=v_{x1}\left(k_x+\frac{eA_{x1}}{\hbar}\right)\nonumber\\
&&~~=b v_{x2}q_{x2}=bv_{x2}\left(k_x+\frac{eA_{x2}}{\hbar}\right),\nonumber\\
&&v_{y1}q_{y1}=v_{y1}\left(k_y+\frac{eA_{y1}}{\hbar}\right)\nonumber\\
&&~~=b v_{y2}q_{y2}=bv_{y2}\left(k_y+\frac{eA_{y2}}{\hbar}\right),\nonumber\\
&&M_1=b M_2
\label{eq:match}
\end{eqnarray}
are satisfied with $b$ a positive proportionality constant. Then it is easy to show that the wave impedances are matched
such that $\chi_{p1}=\chi_{p2}$ and $\chi_{h1}=\chi_{h2}$ and the interband scattering coefficients
$S_{hp}$ and $S_{ph}$ vanish, while the intraband scattering coefficients (or transmission coefficients)
$S_{pp}$ and $S_{hh}$ become unity.
This phenomenon is an equivalent of the temporal Brewster
effect (or temporal total transmission) for electromagnetic waves in time-varying dielectric media \cite{Engheta2021}. Similarly, when the relationships, Eq.~(\ref{eq:match}),
are satisfied with $b<0$, the wave impedances are cross-matched such that
$\chi_{p1}=\chi_{h2}$ and $\chi_{h1}=\chi_{p2}$ and the intraband scattering coefficients $S_{pp}$ and $S_{hh}$ vanish,
while the interband scattering coefficients $S_{hp}$ and $S_{ph}$ become unity.
This can be called temporal total interband transition between the $p$ and $h$ bands.

The Fermi velocity components $v_{xj}$ and $v_{yj}$ are always positive and the wave-vector components $k_x$ and $k_y$ are constants of the motion. Therefore the sign of $b$ can become negative and
the temporal total interband transition can occur only when the time-varying vector potential is present and at
the same time the mass changes its sign (unless both $M_1$ and $M_2$ are zero).
We also notice that if both $\bf k$ and $\bf A$ are along either the $x$ or the $y$ axis
before and after the temporal interface and if the mass is always zero,
then either the temporal Brewster effect or the temporal total interband transition occurs regardless of how the Fermi velocity
and the vector potential vary.
More specifically, the temporal Brewster effect occurs if $q_x$ (or $q_y$) does not change its sign,
while the temporal total interband transition occurs if it does.

In the system of Dirac quasiparticles, the vector potential can be easily tuned by varying the external electric field ${\bf E}(t)$,
which satisfies
\begin{eqnarray}
{\bf E}(t)=-\frac{\partial}{\partial t}{\bf A}(t).
\end{eqnarray}
For instance, the vector potential $A_x$ can be changed suddenly by inducing a sharp spike of electric field in the $x$ direction.
If the electric field is approximated by $E_x(t)=J\delta(t-t_0)$, the change of $A_x$ is given by $A_{x2}-A_{x1}=-J$.
Due to the gauge symmetry, it is always possible to choose the initial vector potential $A_{x1}$ to be zero. Then $A_{x2}$ is given by $-J$.

From the equation of continuity obtained from the Dirac equation, we find that the probability density $\rho$ ($=\vert\Psi\vert^2=\vert\psi_1\vert^2
+\vert\psi_2\vert^2$) in a spatially uniform medium is a constant independent of time. Then it is straightforward to show that
\begin{eqnarray}
T_{pp}+T_{hp}=1,~~T_{hh}+T_{ph}=1,
\label{eq:q1}
\end{eqnarray}
where the transmittances $T_{pp}$ and $T_{hh}$ and the interband transition rates $T_{hp}$ and $T_{ph}$ are defined by
\begin{eqnarray}
&&T_{pp}=\frac{C_{p2}}{C_{p1}}\vert S_{pp}\vert^2,~~T_{hp}=\frac{C_{h2}}{C_{p1}}\vert S_{hp}\vert^2,\nonumber\\
&&T_{hh}=\frac{C_{h2}}{C_{h1}}\vert S_{hh}\vert^2,~~T_{ph}=\frac{C_{p2}}{C_{h1}}\vert S_{ph}\vert^2,
\label{eq:tr1}
\end{eqnarray}
and
\begin{eqnarray}
&&C_{pj}=1+\vert \chi_{pj}\vert^2=1+\frac{\left(\Omega_j-\mu_j\right)^2}{{\vert\nu_j\vert}^2},\nonumber\\
&&C_{hj}=1+\vert \chi_{hj}\vert^2=1+\frac{\left(\Omega_j+\mu_j\right)^2}{{\vert\nu_j\vert}^2}~(j=1,2).
\label{eq:q2}
\end{eqnarray}
Furthermore, we can explicitly show that
\begin{eqnarray}
&&T_{pp}=T_{hh}=\frac{1}{2}\left(1+f\right),\nonumber\\
&&T_{hp}=T_{ph}=\frac{1}{2}\left(1-f\right),
\label{eq:tti}
\end{eqnarray}
where
\begin{eqnarray}
f=\frac{\mu_1\mu_2+v_{x1}v_{x2}q_{x1}q_{x2}+v_{y1}v_{y2}q_{y1}q_{y2}}{\Omega_1\Omega_2}.
\label{eq:f}
\end{eqnarray}
When the wave impedances are matched, $f$ is equal to 1 and $T_{hp}$ and $T_{ph}$ vanish,
while $T_{pp}=T_{hh}=1$.
In contrast, when the impedances are cross-matched, $f$ is equal to $-1$ and $T_{pp}$ and $T_{hh}$
vanish, while $T_{hp}=T_{ph}=1$.

In Fig.~\ref{fig3}, we show the transmittances and the interband transition rates
for a temporal interface as functions of the mass ratio $M_2/M_1$. The common values of the parameters are $k_y=A_{y1}=A_{y2}=0$ and
$M_1/(\hbar k_x v_{x1})=1$. In Fig.~\ref{fig3}(a), we take $v_{x2}q_{x2}=2v_{x1}q_{x1}$. Then the temporal Brewster effect for which
$T_{pp}=T_{hh}=1$ and $T_{hp}=T_{ph}=0$ arises when $M_2/M_1=2$. In Fig.~\ref{fig3}(b), we take $v_{x2}q_{x2}=-2v_{x1}q_{x1}$. Then the temporal total interband transition for which $T_{pp}=T_{hh}=0$ and $T_{hp}=T_{ph}=1$ arises when $M_2/M_1=-2$. The variation of $v_xq_x$ can be achieved by varying $v_x$ or $A_x$. For example, when $v_{x2}=v_{x1}$ and $A_{x1}=0$, we can satisfy the conditions $v_{x2}q_{x2}=2v_{x1}q_{x1}$ and $v_{x2}q_{x2}=-2v_{x1}q_{x1}$ if we choose
$a_{x2}~[\equiv eA_{x2}/(\hbar k_x)]=1$ and $-3$, respectively.

Thus far, we have assumed an abrupt temporal interface characterized by instantaneous and discontinuous
temporal variation. We now consider a more realistic situation in which the temporal variation takes place continuously within a finite interval.
In Fig.~\ref{ff4}, we compare the transmittance at the temporal interface corresponding to
Fig.~\ref{fig3}(a) to those
obtained for continuous interfaces with various values of the interval $\tau_t$. The numerical calculation
of continuous interfaces has been conducted using the well-known invariant imbedding method \cite{SKim2016}.
We assume that $k_y=A_{y}=0$ and $v_x$ is constant. Before $t=0$, the mass energy is $M=M_1=\hbar k_x v_{x}$
and the vector potential $A_x$ is zero. During the interval $0<t<\tau_t$, $M$
and the normalized vector potential $a_{x}$ $[= eA_{x}/(\hbar k_x)]$ change according to
$M=M_1+(M_2-M_1)t/\tau_t$ and $a_x=(t/\tau_t)^2$, respectively. After $t=\tau_t$, they remain to be constants $M=M_2$ and $a_x=1$. With increasing $\tau_t$, the maximum transmittance shifts towards smaller $M_2/M_1$ ratios, while remaining close to 1. For example, when $k_x v_{x}\tau_t=1$, the maximum transmittance is approximately 0.995 and occurs at $M_2/M_1\approx 1.65$. We observe that as long as the quantity $k_x v_{x}\tau_t$, which is proportional to the ratio of $\tau_t$ to the wave period, is not excessively large, the temporal Brewster effect persists in the presence of continuous temporal interfaces.

It is more pertinent to experiments to consider the propagation of wave pulses instead of plane waves. In Fig.~\ref{ff5}, we
consider the propagation of a Gaussian pulse through an abrupt temporal interface at $t=t_0$.
We assume that the mass energy $M$ and the vector potential $A_x$
change discontinuously at $t=t_0$, while $k_y=A_{y}=0$ and $v_x$ remains constant. At $t=0$, the initial pulse consisting of $p$-band states is positioned at $x=0$ and propagates towards the $+x$ direction.
The Gaussian pulse is defined by
\begin{eqnarray}
 u(x,t)=\int_{-\infty}^{\infty} D(k)\Psi(k) dk,
\end{eqnarray}
where
\begin{widetext}
\begin{eqnarray}
&&\Psi(k)=\left\{\begin{array}{l l}
\begin{pmatrix} 1\\ \chi_{p1}(k)\end{pmatrix} e^{i(kx-\omega_{p1} t)}, & \quad \mbox{if $t<t_0$ }\\
e^{i(kx-\omega_{p1}t_0)}\left[S_{pp}(k)\begin{pmatrix} 1\\ \chi_{p2}(k)\end{pmatrix} e^{-i\omega_{p2} (t-t_0)}+S_{hp}(k)\begin{pmatrix} 1\\ \chi_{h2}(k)\end{pmatrix} e^{-i\omega_{h2} (t-t_0)}\right], & \quad \mbox{if $t\ge t_0$ }
\end{array}\right.,\nonumber\\
&&  D(k)=\frac{1}{\sqrt{2\pi}\sigma_k}e^{-\frac{\left(k-k_c\right)^2}{2{\sigma_k}^2}}.
\end{eqnarray}
\end{widetext}
The probability density plotted in Fig.~\ref{ff5} is obtained from
\begin{eqnarray}
 P(x,t)=\frac{\left\vert u(x,t)\right\vert^2}{\int_{-\infty}^{\infty}\left\vert u(x,0)\right\vert^2 dx}.
\end{eqnarray}
The parameter $k_c$ denotes the central wave number and $\sigma_k$ is a measure of the pulse width.
In the present calculation, we set $\sigma_k$ to a fixed value of $0.1 k_c$.
It should be noted that
the frequencies $\omega_{p1}$, $\omega_{p2}$, and $\omega_{h2}$ are also dependent on $k$.
In the cases shown in Figs.~\ref{ff5}(a), \ref{ff5}(b), and \ref{ff5}(c), $M$ is given by $M_1=\hbar k_c v_{x}$ and $A_x$ is zero
at the initial time $t=0$.
The pulse encounters a temporal interface at $t=t_0=5\tau_0$, where the time unit $\tau_0$ is
defined as $\tau_0=1/(\sigma_k v_x)$. Prior to reaching the temporal interface, the group velocity is equal to $v_x/\sqrt{2}$.

We examine four different cases
that correspond to the temporal Brewster effect, partial reflection, total interband transition, and nearly total reflection.
In Fig.~\ref{ff5}(a), which represents the temporal Brewster effect, the mass transitions from $M_1$ to $M_2=2M_1$, and the normalized vector potential $a_{x}$ shifts from zero to one at the interface. In this case, the pulse remains unaffected by the interface and continues to propagate in the same direction with an unchanged group velocity.
In Fig.~\ref{ff5}(b), where $M$ and $a_x$ change to $-2M_1$ and 1 at the interface,
the pulse undergoes a division, propagating simultaneously in both the forward and backward directions. The $p$-band state pulse, moving in the $+x$ direction, advances with a group velocity of $v_{x}/\sqrt{2}$, while the $h$-band state pulse, traveling in the $-x$ direction, possesses a group velocity of $-v_{x}/\sqrt{2}$.
In Fig.~\ref{ff5}(c), which corresponds to the temporal total interband transition, the mass changes from $M_1$ to $M_2=-2M_1$, and $a_{x}$ shifts from zero to $-3$ at the interface. In this case, the states comprising the pulse undergo a complete transition from the $p$-band to the $h$-band. Nevertheless, the pulse continues to propagate in the same direction with an unchanged group velocity. This occurs because the changes in $M$ and $A_x$ result in the interchange of group velocities between the $p$ and $h$ bands. In other words, the group velocity for the $h$-band state after the temporal interface becomes the same as that for the $p$-band state before the interface.
Finally, in Fig.~\ref{ff5}(d), the mass changes from $5\hbar k_{c} v_{x}$ to $-5\hbar k_{c} v_{x}$, while the vector potential remains zero at the interface. As a result of the significant change in mass, an almost complete reflection occurs. The parameter $f$ in Eq.~(\ref{eq:f}) is equal to $-12/13$ and about $96.1 \%$ of the pulse is reflected in the opposite direction.

\subsection{Change of the total energy}

In a time-varying environment, the total energy $\cal E$ is not conserved but varies with time.
We can calculate $\cal E$ using
\begin{eqnarray}
{\cal E}=\frac{\Psi^\dagger{\cal H}\Psi}{\vert \Psi\vert^2}.
\end{eqnarray}
The wave function at an arbitrary time $t$ can be written as
\begin{eqnarray}
\Psi=c\Psi_p+d\Psi_h,
\end{eqnarray}
where $c$ and $d$ are constants and
\begin{eqnarray}
\Psi_p=\begin{pmatrix} 1\\ \chi_{p}\end{pmatrix} e^{-i\omega_{p} t},~\Psi_h=\begin{pmatrix} 1\\ \chi_{h}\end{pmatrix} e^{-i\omega_{h} t}.
\end{eqnarray}
Using
\begin{eqnarray}
{\cal H}\Psi_p=\hbar \omega_p\Psi_p,~{\cal H}\Psi_h=\hbar \omega_h\Psi_h
\end{eqnarray}
and
\begin{eqnarray}
\Psi_p^\dagger\Psi_h=\Psi_h^\dagger\Psi_p=0,
\end{eqnarray}
which follows from $1+\chi_p^*\chi_h=1+\chi_h^*\chi_p=0$,
we can express $\cal E$ as
\begin{eqnarray}
{\cal E}=\frac{\hbar\omega_p\vert c\vert^2\vert\Psi_p\vert^2+\hbar\omega_h\vert d\vert^2\vert\Psi_h\vert^2}{\vert c\vert^2\vert\Psi_p\vert^2+\vert d\vert^2\vert\Psi_h\vert^2}.
\end{eqnarray}
Finally, using Eqs.~(\ref{eq:q1}), (\ref{eq:tr1}), and (\ref{eq:q2}), we straightforwardly
obtain the following expressions for the total wave energy after the temporal interface at $t=t_0$:
\begin{eqnarray}
{\cal E}(t>t_0)=\left\{\begin{array}{l l}
\hbar\omega_{p2}T_{pp}+\hbar\omega_{h2}T_{hp}, & \quad \mbox{$p$ band  }\\
\hbar\omega_{h2}T_{hh}+\hbar\omega_{p2}T_{ph}, & \quad \mbox{$h$ band  }
\end{array}\right.,
\end{eqnarray}
when the states before the temporal interface are $p$- and $h$-band states, respectively.
The change of $\cal E$ is given by
\begin{eqnarray}
\Delta{\cal E}=\left\{\begin{array}{l l}
\hbar\omega_{p2}T_{pp}+\hbar\omega_{h2}T_{hp}-\hbar\omega_{p1}, & \quad \mbox{$p$ band  }\\
\hbar\omega_{h2}T_{hh}+\hbar\omega_{p2}T_{ph}-\hbar\omega_{h1}, & \quad \mbox{$h$ band  }
\end{array}\right..
\end{eqnarray}

\begin{figure}
\includegraphics[width=8.6cm]{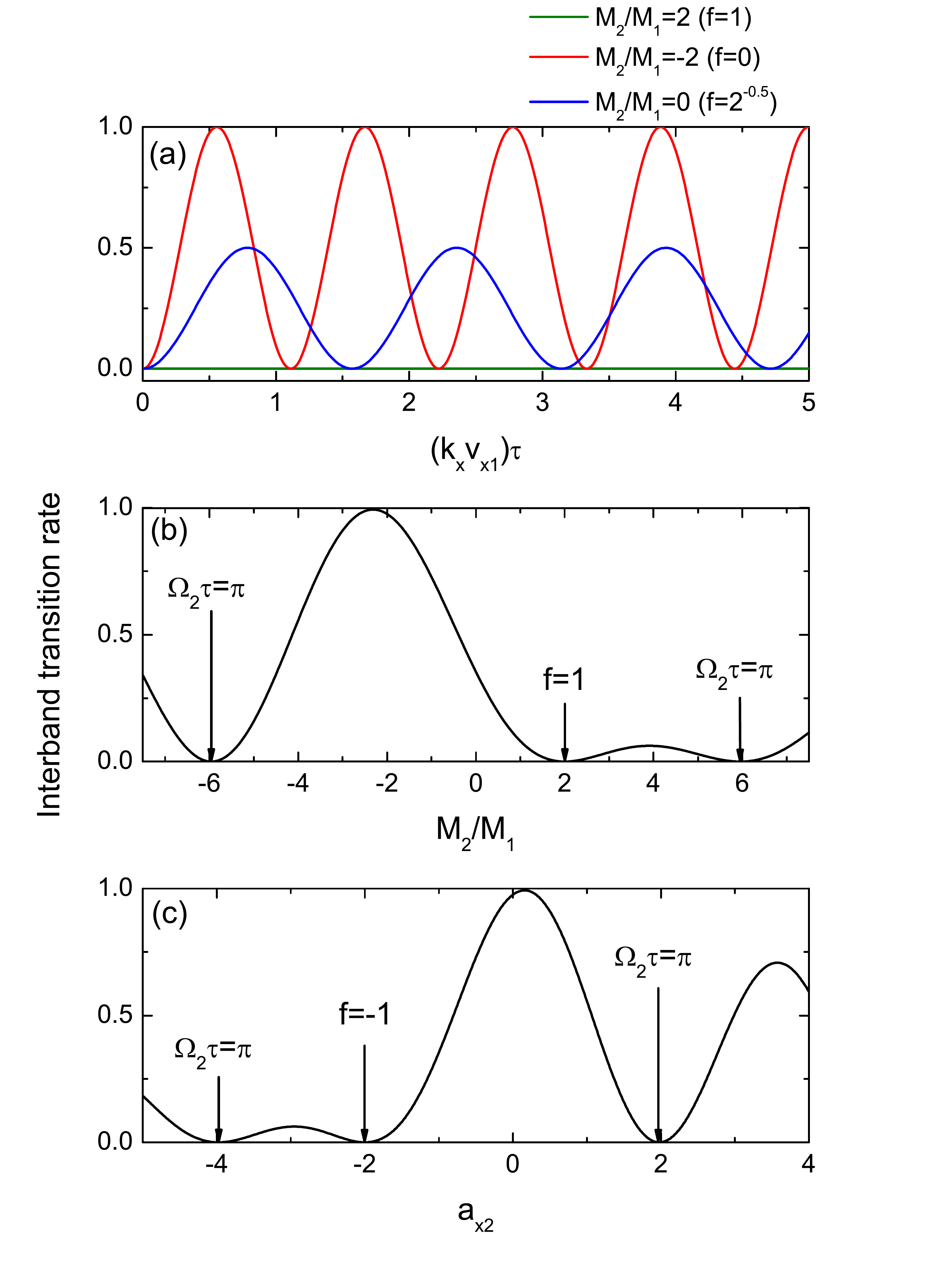}
\caption{Interband transition rate $\tilde T_{hp}$ ($=\tilde T_{ph}$) for a temporal slab of interval $\tau$
plotted versus (a) normalized interval $(k_xv_{x1})\tau$, (b) mass ratio $M_2/M_1$, and (c) normalized
vector potential $a_{x2}$ $[=eA_{x2}/(\hbar k_x)]$. In (a), we set $k_y={\bf A}_{1}={\bf A}_{2}=0$,
$M_1/(\hbar k_x v_{x1})=1$, $v_{x2}/v_{x1}=2$, and $M_2/M_1=2$, $-2$, 0.
For $M_2/M_1=2$, $f$ is equal to 1 and $\tilde T_{hp}=0$.
For $M_2/M_1=-2$, $f$ is equal to 0 and $\tilde T_{hp}=\sin^2\left(2\sqrt{2}k_xv_{x1}\tau\right)$.
For $M_2/M_1=0$, $f$ is equal to $1/\sqrt{2}$ and $\tilde T_{hp}=0.5\sin^2\left(2k_xv_{x1}\tau\right)$. In (b), we set $k_y={\bf A}_{1}={\bf A}_{2}=0$,
$M_1/(\hbar k_x v_{x1})=1$, $v_{x2}/v_{x1}=2$, and $k_xv_{x1}\tau=0.5$. The interband transition rate vanishes
for either $f=1$ corresponding to $M_2/M_1=2$ or $\Omega_2\tau=\pi$ corresponding to $M_2/M_1\approx\pm 5.956$. In (c), we set $k_y=A_{x1}=A_{y1}=A_{y2}=0$,
$M_1/(\hbar k_x v_{x1})=1$, $M_2/M_1=-2$, $v_{x2}/v_{x1}=2$, and $k_xv_{x1}\tau=0.5$.
The interband transition rate vanishes
for either $f=-1$ corresponding to $M_2/M_1\approx -1.998$ or $\Omega_2\tau=\pi$ corresponding to $M_2/M_1\approx 1.978$ and $-3.978$.}
\label{fig4}
\end{figure}

\section{Temporal slab}
\label{sec4}

In this section, we consider a simple temporal slab of interval $\tau$ such that at $t=t_0$, the parameters change from
$U_1$, $A_{x1}$, $A_{y1}$, $v_{x1}$, $v_{y1}$, $v_{tx1}$, $v_{ty1}$, and $M_1$ to
$U_2$, $A_{x2}$, $A_{y2}$, $v_{x2}$, $v_{y2}$, $v_{tx2}$, $v_{ty2}$, and $M_2$, and then later at $t=t_0+\tau$,
they change back to their initial values.
A schematic of the temporal scattering process through a temporal slab is shown in Fig.~\ref{fig1}(b).
Since the temporal scattering proceeds only in the $+t$ direction, we can easily obtain the scattering coefficients
for the temporal slab, which we call $\tilde S_{pp}$, $\tilde S_{hp}$, $\tilde S_{hh}$, and $\tilde S_{ph}$,
in terms of those for the two temporal interfaces:
\begin{eqnarray}
&&\tilde S_{pp}=S_{pp}S_{pp}^\prime e^{-i\omega_{p2}\tau}+S_{hp}S_{ph}^\prime e^{-i\omega_{h2}\tau},\nonumber\\
&&\tilde S_{hp}=S_{pp}S_{hp}^\prime e^{-i\omega_{p2}\tau}+S_{hp}S_{hh}^\prime e^{-i\omega_{h2}\tau},\nonumber\\
&&\tilde S_{hh}=S_{hh}S_{hh}^\prime e^{-i\omega_{h2}\tau}+S_{ph}S_{hp}^\prime e^{-i\omega_{p2}\tau},\nonumber\\
&&\tilde S_{ph}=S_{hh}S_{ph}^\prime e^{-i\omega_{h2}\tau}+S_{ph}S_{pp}^\prime e^{-i\omega_{p2}\tau},
\end{eqnarray}
where the scattering coefficients for the interface at $t=t_0+\tau$, $S_{pp}^\prime$, $S_{hp}^\prime$, $S_{hh}^\prime$, and $S_{ph}^\prime$, are
obtained by exchanging $\Omega_1$, $\mu_1$, and $\nu_1$ with $\Omega_2$, $\mu_2$, and $\nu_2$, respectively,
in the definitions of
$S_{pp}$, $S_{hp}$, $S_{hh}$, and $S_{ph}$ given by Eqs.~(\ref{eq:sc1}) and (\ref{eq:sc2}).
In the present case, the transmittances $\tilde T_{pp}$ and $\tilde T_{hh}$ and the interband transition rates $\tilde T_{hp}$ and $\tilde T_{ph}$ are defined by
\begin{eqnarray}
&&\tilde T_{pp}=\vert \tilde S_{pp}\vert^2,~~\tilde T_{hp}=\frac{C_{h1}}{C_{p1}}\vert \tilde S_{hp}\vert^2,\nonumber\\
&&\tilde T_{hh}=\vert \tilde S_{hh}\vert^2,~~\tilde T_{ph}=\frac{C_{p1}}{C_{h1}}\vert \tilde S_{ph}\vert^2.
\end{eqnarray}
By straightforward calculations, we obtain
\begin{eqnarray}
&&\tilde T_{pp}=\tilde T_{hh}=1-\left(1-f^2\right)
\sin^2\left(\Omega_2\tau\right),\nonumber\\
&&\tilde T_{hp}=\tilde T_{ph}=\left(1-f^2\right)
\sin^2\left(\Omega_2\tau\right).
\label{eq:tts}
\end{eqnarray}
The conservation laws
\begin{eqnarray}
\tilde T_{pp}+\tilde T_{hp}=1,~~\tilde T_{hh}+\tilde T_{ph}=1
\end{eqnarray}
are easily seen to be satisfied.
We note that all of $\tilde T_{pp}$, $\tilde T_{hp}$, $\tilde T_{hh}$, and $\tilde T_{ph}$ depend periodically on the interval $\tau$ with period $\pi/\Omega_2$.
The temporal Brewster effect occurs when $f=\pm 1$ or $\tau=n\pi/\Omega_2$ with $n$ an arbitrary integer,
whereas the temporal total interband transition occurs when $f=0$ and $\tau=(n+1/2)\pi/\Omega_2$ with $n$ an arbitrary integer.
In the case of $f=-1$, the temporal Brewster effect arises due to the two consecutive temporal total interband transitions at $t=0$ and $t=\tau$.
The change of the total energy when the states before the temporal slab are $p$- and $h$-band states is given by
\begin{eqnarray}
\Delta{\cal E}=\left\{\begin{array}{l l}
-2\hbar\Omega_1\left(1-f^2\right)\sin^2\left(\Omega_2\tau\right), & \quad \mbox{$p$ band  }\\
2\hbar\Omega_1\left(1-f^2\right)\sin^2\left(\Omega_2\tau\right), & \quad \mbox{$h$ band  }
\end{array}\right..
\end{eqnarray}

If we ignore the scalar potential and the tilt velocity and set $k_y=A_{y1}=A_{y2}=0$,
we can derive the explicit expressions
\begin{eqnarray}
&&\tilde S_{hp}=i\frac{\left(\mu_1\nu_2-\mu_2\nu_1\right)\left(\Omega_1-\mu_1\right)}{\nu_1\Omega_1\Omega_2}\sin\left(\Omega_2\tau\right),\nonumber\\
&&\tilde S_{ph}=i\frac{\left(\mu_1\nu_2-\mu_2\nu_1\right)\left(\Omega_1+\mu_1\right)}{\nu_1\Omega_1\Omega_2}\sin\left(\Omega_2\tau\right),\nonumber\\
&&\tilde T_{hp}=\tilde T_{ph}=\frac{\left(\mu_1\nu_2-\mu_2\nu_1\right)^2}{{\nu_1}^2{\Omega_1}^2{\Omega_2}^2}\sin^2\left(\Omega_2\tau\right).
\end{eqnarray}
In the special case where $M_1=A_{x1}=A_{x2}=0$ and $v_{x1}=v_{x2}=v_F$, they are simplified to
\begin{eqnarray}
&&\tilde S_{hp}=\tilde S_{ph}=-i\frac{\mu_2}{\Omega_2}\sin\left(\Omega_2\tau\right),\nonumber\\
&&\tilde T_{hp}=\tilde T_{ph}=\frac{{\mu_2}^2}{{\Omega_2}^2}\sin^2\left(\Omega_2\tau\right),
\end{eqnarray}
where $\Omega_2=\sqrt{{M_2}^2+(\hbar v_F k_x)^2}/\hbar$.
The expression for the interband transition amplitudes $\tilde S_{hp}$ and $\tilde S_{ph}$ agrees precisely with Eq.~(3) derived in Ref.~\onlinecite{Reck2017}.

In Fig.~\ref{fig4}, we illustrate the dependencies of the interband transition rates $\tilde T_{hp}$ and $\tilde T_{ph}$
on the normalized interval $(k_xv_{x1})\tau$, mass ratio $M_2/M_1$, and normalized vector potential $a_{x2}$.
In Fig.~\ref{fig4}(a), we set $k_y={\bf A}_{1}={\bf A}_{2}=0$,
$M_1/(\hbar k_x v_{x1})=1$, $v_{x2}/v_{x1}=2$, and $M_2/M_1=2$, $-2$, 0.
For $M_2/M_1=2$, $f$ is equal to 1 and the interband transition rate vanishes.
For $M_2/M_1=-2$, $f$ is equal to 0 and the interband transition rate is a periodic function given by $\tilde T_{hp}=\sin^2\left(2\sqrt{2}k_xv_{x1}\tau\right)$.
For $M_2/M_1=0$, $f$ is equal to $1/\sqrt{2}$ and $\tilde T_{hp}=0.5\sin^2\left(2k_xv_{x1}\tau\right)$. In Fig.~\ref{fig4}(b), we set $k_y={\bf A}_{1}={\bf A}_{2}=0$,
$M_1/(\hbar k_x v_{x1})=1$, $v_{x2}/v_{x1}=2$, and $k_xv_{x1}\tau=0.5$. Then the interband transition rate vanishes
for either $f=1$ corresponding to $M_2/M_1=2$ or $\Omega_2\tau=\pi$ corresponding to $M_2/M_1\approx\pm 5.956$. In Fig.~\ref{fig4}(c), we set $k_y=A_{x1}=A_{y1}=A_{y2}=0$,
$M_1/(\hbar k_x v_{x1})=1$, $M_2/M_1=-2$, $v_{x2}/v_{x1}=2$, and $k_xv_{x1}\tau=0.5$.
Then the interband transition rate vanishes
for either $f=-1$ corresponding to $M_2/M_1\approx -1.998$ or $\Omega_2\tau=\pi$ corresponding to $M_2/M_1\approx 1.978$ and $-3.978$.

\begin{figure}
\includegraphics[width=8.6cm]{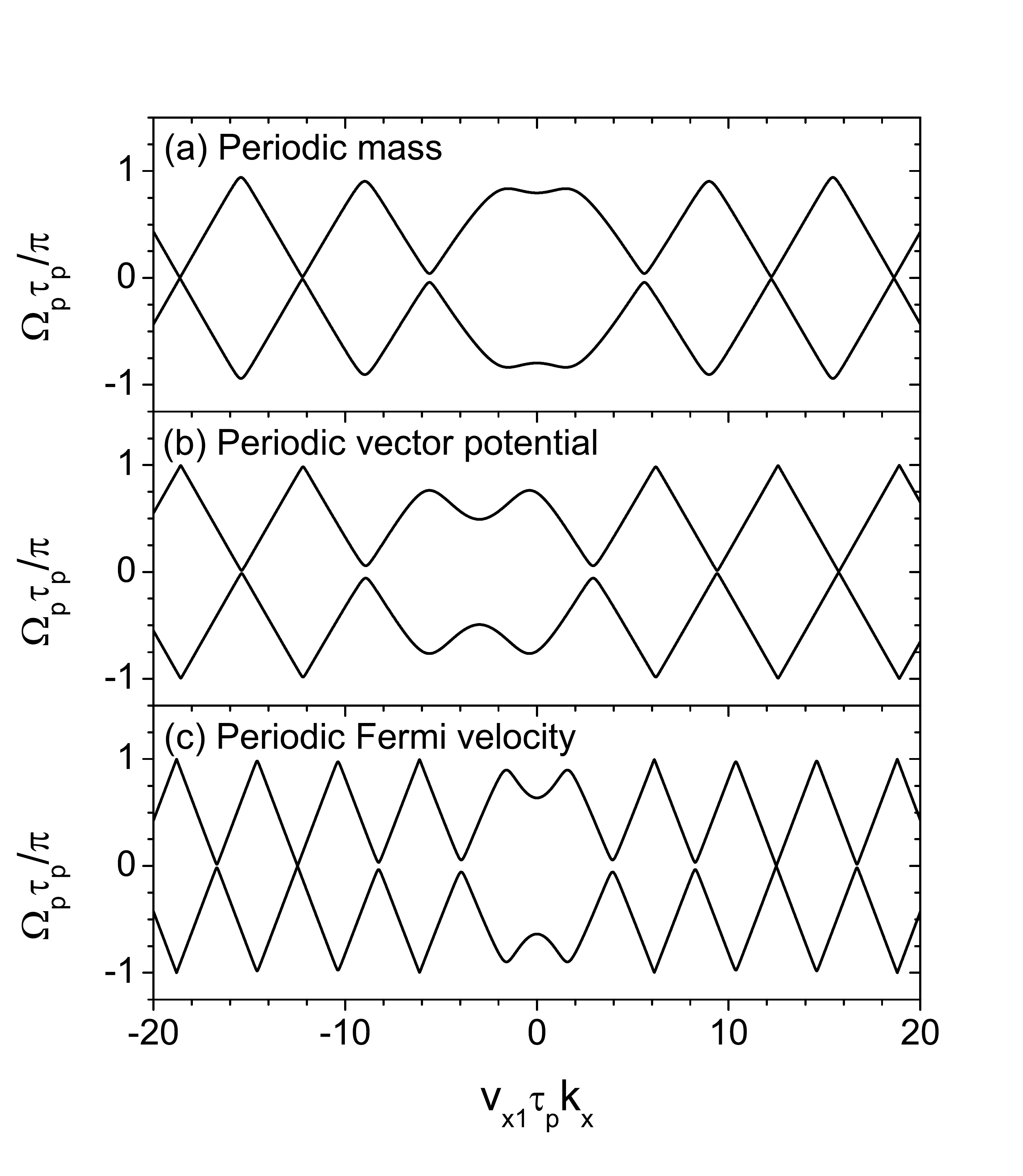}
\caption{Normalized frequency $\Omega_p\tau_p/\pi$ plotted versus normalized wavenumber $(v_{x1}\tau_p) k_x$
for bilayer Dirac temporal crystals where (a) the normalized mass $\tilde m=M\tau_p/\hbar$ alternates between 1 and 4 while $k_y={\bf A}_1={\bf A}_2=0$ and $v_{x1}=v_{x2}$,
(b) the normalized vector potential $v_{x1}eA_x\tau_p/\hbar$ alternates between 1 and 5 while $k_y=A_{y1}=A_{y2}=0$,
$\tilde m_1=\tilde m_2=2$, and $v_{x1}=v_{x2}$, and (c) the Fermi velocity $v_x$ alternates between $v_{x1}$ and
$2v_{x1}$ while $k_y={\bf A}_1={\bf A}_2=0$ and $\tilde m_1=\tilde m_2=2$. Only the curves in one period in the range $-1<\Omega_p\tau_p/\pi<1$ are displayed. These curves are repeated periodically with period 2 along the $\Omega_p\tau_p/\pi$ axis.}
\label{fig5}
\end{figure}

\section{Absence of momentum gaps in Dirac temporal crystals}
\label{sec5}

In the previous sections, we have shown that the temporal scattering for Dirac particles and waves can be caused by the temporal variation of
the mass, Fermi velocity, and vector potential. In this section, we consider a bilayer temporal crystal where these quantities vary periodically as a function of time with period $\tau_p$ such that within one period
\begin{eqnarray}
 &&\left(M(t),A_x(t),A_y(t),v_x(t),v_y(t)\right)\nonumber\\
 &&=\begin{cases}
           \left(M_1,A_{x1},A_{y1},v_{x1},v_{y1}\right), & \mbox{if $0<t<t_1$}\\
           \left(M_2,A_{x2},A_{y2},v_{x2},v_{y2}\right), & \mbox{if $t_1<t<\tau_p$}
         \end{cases},
\end{eqnarray}
where $\tau_p=t_1+t_2$. We assume that the scalar potential and the tilt velocity are zero because they do not cause any temporal scattering.
From the continuity of the wave function at $t=t_1$ and $t=\tau_p$ and the Floquet-Bloch theorem
demanding
\begin{eqnarray}
 \psi\left(t+\tau_p\right)=e^{-i{\Omega}_p \tau_p}\psi\left(t\right),
\end{eqnarray}
we obtain the characteristic matrix
\begin{widetext}
\begin{eqnarray}
R =\begin{pmatrix}
e^{-i\Omega_1t_1} & e^{i\Omega_1t_1} & -e^{-i\Omega_2t_1} & -e^{i\Omega_2t_1}\\
\chi_{p1}e^{-i\Omega_1t_1} & \chi_{h1}e^{i\Omega_1t_1} & -\chi_{p2}e^{-i\Omega_2t_1} & -\chi_{h2}e^{i\Omega_2t_1}\\
e^{-i\Omega_p\tau_p} & e^{-i\Omega_p\tau_p} & -e^{-i\Omega_2\tau_p} & -e^{i\Omega_2\tau_p}\\
\chi_{p1}e^{-i\Omega_p\tau_p} & \chi_{h1}e^{-i\Omega_p\tau_p} & -\chi_{p2}e^{-i\Omega_2\tau_p} & -\chi_{h2}e^{i\Omega_2\tau_p}
\end{pmatrix}.
\end{eqnarray}
\end{widetext}
The frequency $\Omega_p$ is the eigenfrequency of the temporal crystal. The momentum gap would correspond to the region of the momentum
(or wave vector) in
which $\Omega_p$ has no real-valued solution.
The dispersion relation of the temporal crystal is obtained from the condition that the determinant of $R$ is zero
and takes the simple form
\begin{eqnarray}
&&\cos{\left({\Omega}_p \tau_p\right)}=\cos{\left(\Omega_1 t_1\right)}
  \cos{\left(\Omega_2 t_2\right)}\nonumber\\
&&~~~~~~~~~~~~~~~-f \sin{\left(\Omega_1 t_1\right)}
  \sin{\left(\Omega_2 t_2\right)}\nonumber\\
  &&=F\cos{\left(\Omega_1 t_1+\Omega_2 t_2\right)}
  +G\cos{\left(\Omega_1 t_1-\Omega_2 t_2\right)},
\end{eqnarray}
where $f$ is defined by Eq.~(\ref{eq:f}) and
\begin{eqnarray}
F+G=1,~~F-G=f.
\end{eqnarray}
From this, we notice that $\vert\cos{\left(\Omega_p \tau_p\right)}\vert$ satisfies the inequality
\begin{eqnarray}
\vert\cos{\left(\Omega_p \tau_p\right)}\vert \le {\rm Max}\left(1,\vert f\vert\right).
\end{eqnarray}
From the explicit form of $f$, we can prove that
\begin{eqnarray}
&&\left(1-f^2\right)\left(\Omega_1\Omega_2\right)^2=
\left(\mu_1 v_{x2}q_{x2}-\mu_2 v_{x1}q_{x1}\right)^2\nonumber\\&&~~~~~+\left(\mu_1 v_{y2}q_{y2}-\mu_2 v_{y1}q_{y1}\right)^2\nonumber\\&& ~~~~~+\left(v_{x1}q_{x1}v_{y2}q_{y2}-v_{x2}q_{x2}v_{y1}q_{y1}\right)^2 \geq 0.
\end{eqnarray}
Therefore $\vert f\vert$ is not larger than 1 and the dispersion relation always has a real solution.
We conclude that there never appears a momentum gap in bilayer Dirac temporal crystals
in a sharp contrast to the case of electromagnetic waves.

In Fig.~\ref{fig5}, we show the dispersion relations for bilayer Dirac temporal crystals when the mass, vector potential, or Fermi velocity varies periodically with period $\tau_p$.
In Fig.~\ref{fig5}(a), the normalized mass $\tilde m=M\tau_p/\hbar$ alternates between 1 and 4 while $k_y={\bf A}_1={\bf A}_2=0$ and $v_{x1}=v_{x2}$.
In Fig.~\ref{fig5}(b), the normalized vector potential $v_{x1}eA_x\tau_p/\hbar$ alternates between 1 and 5 while $k_y=A_{y1}=A_{y2}=0$,
$\tilde m_1=\tilde m_2=2$, and $v_{x1}=v_{x2}$. In Fig.~\ref{fig5}(c), the Fermi velocity $v_x$ alternates between $v_{x1}$ and
$2v_{x1}$ while $k_y={\bf A}_1={\bf A}_2=0$ and $\tilde m_1=\tilde m_2=2$. These curves are repeated periodically with period 2 along the $\Omega_p\tau_p/\pi$ axis.
In all the cases, we find that there appears no momentum gap. The influence of the periodic variation is seen to be strongest in the region where $\vert k_x\vert$ is small
and becomes weaker as it increases.

\section{Discussion and Conclusion}
\label{sec6}

Let us discuss briefly the experimental feasibility of the effects explored in this paper.
In 2D Dirac materials, the quantities such as the scalar and vector potentials, Fermi velocity, tilt velocity, and mass can be tuned
readily by various means.
It is easy to tune the scalar potential by applying a uniform gate voltage to the whole layer \cite{Kats2006}.
A spatially uniform and time-dependent vector potential can be most easily generated by applying a uniform electric field parallel to the 2D layer \cite{Men2021}.
It has been suggested that the mass, or the band gap between the upper and lower Dirac cones, in 2D materials such as silicene and germanene can be varied by tuning
the electric field applied perpendicularly to the layer \cite{Ni2012,Gho2020}.
It has also been proposed that the Fermi velocity of the systems such as graphene nanoribbons and carbon nanotubes can be tuned by applying
a uniform electric field across such materials \cite{Fer2017}.
The tilt velocity of the $8Pmmn$ borophene sheet that has tilted Dirac cones has been proposed to be tunable by applying an electric field perpendicular to the sheet \cite{Jafari2019}.
There exist other systems exhibiting Dirac cones in their energy dispersion such as suitably designed photonic crystals and metamaterials and cold atoms
trapped in optical lattices.
It is also possible to vary the parameters of those systems temporally by various means.
For example, a method to tune the effective vector potential for polaritons
supported by a strained honeycomb metasurface composed of interacting dipole emitters/antennas has been proposed \cite{Mann2020}.
The variation of the parameters of cold atomic systems in optical lattices should also be possible by optical means.

The temporal scattering effects will be manifested experimentally in various
physical quantities. In electronic Dirac materials, the temporal reflection of electron matter waves
is expected to cause a modification in electronic currents. In quasi-one-dimensional mesoscopic systems, it has been well-known that the conductance is directly proportional to the transmittance according to the Landauer formula. Therefore, through the measurement of current and conductance in the presence of temporal variations, it is feasible to experimentally investigate the effects of temporal reflection. In photonic metamaterials
exhibiting Dirac-type dispersion, it should be possible to study directly the propagation of electromagnetic
wave pulses similar to those considered in Fig.~\ref{ff5}. In all of these cases, the observability can be assessed by evaluating the relative change in wave transmittance caused by temporal variations.
Let us suppose that only the mass energy or the gap between the two Dirac cones varies from zero to a nonzero value. To achieve a temporal reflectance of 0.05 at a temporal interface, it is necessary to set $f=0.9$ in Eq.~(\ref{eq:tti}), which can be realized by opening a gap of $M=0.48~ \hbar k_x v_x$. Similarly,
to achieve a maximum temporal reflectance of 0.05 across a temporal slab, it is necessary to set $f^2=0.95$ in Eq.~(\ref{eq:tts}), which can be accomplished by opening a gap of $M=0.23~ \hbar k_x v_x$. These values are relatively small and can be easily achieved using the experimental techniques described in Refs.~\onlinecite{Ni2012,Gho2020}.

In conclusion, we have studied the influence of the temporal variations of the medium parameters on the propagation of Dirac-type waves in
materials where the quasiparticles are described by a generalized version of the pseudospin-1/2 Dirac equation.
We have derived the scattering coefficients associated with the temporal interfaces and slabs analytically
and found that the temporal scattering is caused by the changes of the mass, Fermi velocity, and vector potential,
but does not arise from the changes
of the scalar potential and tilt velocity.
Using the analytical expressions for the temporal transmittances and the interband transition rates, we have obtained
the explicit conditions for which the temporal Brewster effect and total interband transition occur.
We have also proved that in bilayer Dirac temporal crystals where the parameters alternate between two different values, momentum gaps do not appear in a sharp contrast to the classical waves.
It is highly desirable to generalize the present investigation to the case where the parameters change arbitrarily in time. This will be a subject of research in the future.

\acknowledgments
This research was supported through a National Research
Foundation of Korea Grant (NRF-2022R1F1A1074463)
funded by the Korean Government.
It was also supported by the Basic Science Research Program through the National Research Foundation of Korea funded by the Ministry of Education (NRF-2021R1A6A1A10044950)
and by the Global Frontier Program (2014M3A6B3063708).

\bibliography{timer_r}

\end{document}